\documentclass[aps,prl,groupedaddress,preprint,showpacs]{revtex4}
\usepackage{graphicx,graphics,psfrag,amsmath,calc,amssymb}

\begin{document}
\title{Coexistence of Triplet Superconductivity and Spin Density Wave}
\author{Wei Zhang}
\author{C. A. R. S\'{a} de Melo}
\affiliation{School of Physics, Georgia Institute of Technology, 
Atlanta, Georgia 30332}

\date{\today}

\begin{abstract}
We discuss the possibility of coexistence 
of spin density wave 
(antiferromagnetism) and triplet superconductivity 
as a particular example of a broad class 
of systems where the interplay of magnetism and superconductivity 
is important. We focus on the case of quasi-one-dimensional metals,
where it is known that antiferromagnetism
is in close proximity to triplet superconductivity in the pressure
versus temperature phase diagram. Over a range of pressures, we propose 
an intermediate non-uniform phase consisting of 
antiferromagnetic and triplet superconducting orders. 
In the coexistence region, we propose a flop transition in the spin density
wave order parameter vector, which affects the nature 
of the superconducting state.

\end{abstract}

\maketitle

%

The competition or coexistence of magnetic order and superconductivity 
is a very important problem in condensed matter physics. 
There is a broad class of systems that present magnetic order 
and superconductivity in close vicinity. 
One of the most important systems are the Copper oxides, 
where singlet superconductivity is found next 
to antiferromagnetism~\cite{dagotto-94}. 
Another interesting system is Strontium Ruthenate ${\rm Sr_2RuO_4}$, 
where the proximity to ferromagnetism 
has been argued as being important to the existence of 
possible triplet superconductivity in these materials~\cite{maeno-94}. 
Furthermore the ferromagnetic 
superconductors ${\rm ZrZn_2}$ and ${\rm UGe_2}$ have stimulated 
a debate on the coexistence of ferromagnetism and triplet 
or singlet superconductivity~\cite{lonzarich-01,saxena-00}. 
However, unlike any of these previous examples, 
we discuss in this manuscript a 
quasi-one-dimensional organic superconductor, 
the Bechgaard salt ${\rm (TMTSF)_2PF_6}$, which has a phase diagram 
of neighboring antiferromagnetism (AFM) and 
triplet superconductivity (TSC)~\cite{ishiguro-98}. 

The antiferromagnetic state of ${\rm (TMTSF)_2PF_6}$ 
is present at temperatures $T<12 $ K and pressures $P <$ 6 kbar, 
and is characterized by a spin density wave (SDW)~\cite{ishiguro-98}. 
The SDW order parameter ${\bf N}$ (Neel vector) is anisotropic, 
having an easy axis along the crystallographic 
${\bf b}^\prime$ axis~\cite{mortensen-81}, which is also 
the intermediate direction for conductivity. 
This antiferromagnetic state is supressed 
at pressures higher than 6 kbar, where a superconducting 
instability takes over at low temperatures ($T <T_c \approx 1.2$ K). 
This superconducting state is very likely to be 
triplet, as suggested by upper critical fields~\cite{lee-97} 
and NMR~\cite{lee-03} measurements. 
Recent experiments~\cite{vuletic-02, kornilov-04, lee-05} 
suggest a region of macroscopic coexistence between TSC and SDW, 
where both orders are non-uniform. 
This coexistence region can be related to 
existing theoretical proposals. 
For instance, strictly one-dimensional theories 
invoking SO(4) symmetry~\cite{podolsky-04}, or negative 
interface energies~\cite{zhang-05} have allowed for 
coexisting TSC and SDW. 
However, these previous theories are not directly applicable
to three-dimensional but highly anisotropic superconductors
like the Bechgaard salts, where the SO(4) symmetry is absent, 
and negative interface energies are not necessary conditions 
for the coexistence.
In this manuscript, we derive microscopically the pressure versus 
temperature phase diagram indicating the TSC, the SDW and the TSC/SDW
phases, and show that the TSC and SDW order parameters 
are both non-uniform in the coexistence region.
Furthermore, we investigate the effects of an external
magnetic field and suggest that a canting transition of the 
SDW order parameter may occur, and alter the nature of 
the TSC state in the coexistence region.

The compound ${\rm (TMTSF)_2PF_6}$ 
can be described approximately by an orthorhombic lattice with dispersion
\begin{equation}
\label{eqn:dispersion}
\epsilon_{\bf k} = -\vert t_x \vert \cos(k_x a) -\vert t_y \vert \cos(k_y b)
-\vert t_z \vert \cos(k_z c),
\end{equation}
where transfer integrals $\vert t_x \vert$, $\vert t_y \vert$ 
and $\vert t_z \vert$ satisfy the relations 
$\vert t_x \vert \gg \vert t_y \vert \gg \vert t_z \vert$
representing the quasi-one-dimensionality.
Here, $a$, $b$, and $c$ correspond to unit cell lengths 
along the crystallographic axes ${\bf a}$(x), 
${\bf b}^\prime$(y) and ${\bf c}^*$(z), respectively. 

We use natural units ($\hbar = k_B = c =1$) 
and work with Hamiltonian ${\cal H} = {\cal H}_0 +{\cal H}_{\rm int}$, 
where the non-interacting part is 
${\cal H}_0 = \sum_{{\bf k},\alpha} \xi_{\bf k} 
c_{{\bf k},\alpha}^\dagger c_{{\bf k},\alpha}$, 
and $\xi_{\bf k} = \epsilon_{\bf k} - \mu$ is 
the dispersion shifted by the chemical potential, which
may include a Hartree shift.
The interaction part is
\begin{eqnarray}
\label{eqn:Hint}
{\cal H}_{\rm int} &=& \sum_{\bf k k^\prime p} 
\sum_{\alpha\beta\gamma\delta} 
V({\bf k}, {\bf k}^\prime) 
{\bf d}_{\alpha\beta}^\dagger ({\bf k}, {\bf p}) \cdot
{\bf d}_{\gamma\delta} ({\bf k}^\prime, {\bf p})
\nonumber \\
&+&
\sum_{\bf k k^\prime q} 
\sum_{\alpha\beta\gamma\delta} 
J({\bf q})
{\bf s}_{\alpha\beta}^\dagger ({\bf k}, {\bf q}) \cdot
{\bf s}_{\gamma\delta} ({\bf k}^\prime, {\bf q}),
\end{eqnarray}
where the first and second terms describe interactions 
in TSC and SDW channels, respectively. These interactions
allow for the possibility of competition or coexistence of
TSC and SDW instabilities at low temperatures.
Here, $\alpha$, $\beta$, $\gamma$ and $\delta$ are spin indices and 
${\bf k}$, ${\bf k}^\prime$, ${\bf p}$ and ${\bf q}$ represent linear momenta. 
The vector operator 
${\bf d}_{\alpha\beta}^\dagger ({\bf k}, {\bf p}) \equiv 
c_{{\bf k}+{\bf p}/2,\alpha}^\dagger {\bf v}_{\alpha\beta} 
c_{-{\bf k}+{\bf p}/2,\beta}^\dagger$, 
and ${\bf s}_{\alpha\beta}^\dagger ({\bf k}, {\bf q}) \equiv 
c_{{\bf k}-{\bf q}/2,\alpha}^\dagger \textrm{\boldmath $\sigma$}_{\alpha\beta} 
c_{{\bf k}+{\bf q}/2,\beta}$.
The matrix 
${\bf v}_{\alpha\beta}=(i \textrm{\boldmath $\sigma$}\sigma_y)_{\alpha\beta}$, 
and $\sigma_i$ are Pauli matrices. 
In the case of weak spin-orbit coupling, 
the TSC interaction $V({\bf k}, {\bf k}^\prime)$ 
can be choosen as~\cite{duncan-01}
\begin{equation}
\label{eqn:couplingTS}
V({\bf k}, {\bf k}^\prime) = 
V h_\Gamma({\bf k}, {\bf k}^\prime) \phi_\Gamma({\bf k}) 
\phi_\Gamma({\bf k}^\prime)
\end{equation}
where $V$ is a prefactor with dimension of energy, 
$h_\Gamma({\bf k}, {\bf k}^\prime)$ $[\phi_\Gamma({\bf k})]$ 
characterizes the momentum dependence [symmetry basis function]
for an irreducible representation $\Gamma$ 
of the orthorhombic group $D_{\rm 2h}$. Without loss of the 
generality regarding symmetry properties, we take 
$h_\Gamma({\bf k}, {\bf k}^\prime) = 1$, and consider only 
unitary triplet states corresponding to $p_x$ symmetry.

The order parameter for TSC can be defined as
\begin{equation}
\label{eqn:D-vector}
{\bf D}({\bf p}) = \langle \sum_{{\bf k},\alpha\beta}
V \phi_\Gamma({\bf k})
{\bf d}_{\alpha\beta} ({\bf k}, {\bf p}) \rangle,
\end{equation}
while the SDW order parameter can be defined as
\begin{equation}
\label{eqn:Neel}
{\bf N}({\bf q}) = J({\bf q}) \langle \sum_{{\bf k},\alpha\beta} 
{\bf s}_{\alpha\beta} ({\bf k}, {\bf q}) \rangle.
\end{equation}
With these definitions, the effective Hamiltonian is
\begin{equation}
\label{eqn:MFH}
{\cal H}_{\rm eff} = {\cal H}_0 + 
{\cal H}_{\rm TSC} + {\cal H}_{\rm SDW},
\end{equation}
where the TSC contribution is $ {\cal H}_{\rm TSC} = \sum_{\bf p} 
[{\bf D}^\dagger ({\bf p}) \cdot \sum_{{\bf k},\alpha\beta} 
\phi_\Gamma({\bf k}) {\bf d}_{\alpha\beta} ({\bf k}, {\bf p}) + {\rm H.C.}]
- \sum_{\bf p} {\bf D}^\dagger({\bf p}) \cdot {\bf D}({\bf p})/V$, 
and the TSC term is 
${\cal H}_{\rm SDW} =
\sum_{\bf q} [ {\bf N}(-{\bf q}) \cdot \sum_{{\bf k},\alpha\beta} 
{\bf S}_{\alpha\beta}({\bf k}, {\bf q}) + {\rm H.C.}] - 
\sum_{\bf q} {\bf N}(-{\bf q}) \cdot {\bf N}({\bf q})/J({\bf q}) $.
The effective action of this Hamiltonian as a function 
of ${\bf D}({\bf p})$ and ${\bf N}({\bf q})$ 
is obtained by integrating out the fermions. 
The quadratic terms are 
\begin{eqnarray}
\label{eqn:2-order}
S_2^{\rm TSC} &=& \sum_{\bf p} A({\bf p}) {\bf D}^\dagger({\bf p}) 
\cdot {\bf D}({\bf p}), \nonumber \\
S_2^{\rm SDW} &=& \sum_{\bf q} B({\bf q}) {\bf N}(-{\bf q}) 
\cdot {\bf N}({\bf q}),
\end{eqnarray}
with coefficients 
$A({\bf p}) = - 2 V^{-1}
-2 T \sum_{k} 
G({\bf k} + {\bf p}/2, \omega_n) \phi_\Gamma ( {\bf k}+{\bf p}/2 ) 
G(-{\bf k}+{\bf p}/2,-\omega_n ) \phi_\Gamma (-{\bf k}+{\bf p}/2)$, 
and  $B({\bf q}) = -2 J^{-1}({\bf q}) 
- 2 T \sum_{k} G({\bf k}, \omega_n ) 
G({\bf k}+{\bf q}, \omega_n)$. 
Here, $T$ is the temperature, 
$G({\bf k},\omega_n) = 1/(i\omega_n + \xi_{\bf k})$ is the fermion propagator, 
$\omega_n$ is fermionic Matsubara frequency, 
and $\sum_k = \sum_{{\bf k},\omega_n}$ is used to shorthand notation.  

In what follows we make some standard assumptions.
First, we assume that the saddle point TSC order parameter is 
dominated by the zero center of mass momentum component 
${\bf D}_0 \equiv {\bf D}({\bf p}=0)$. 
Second, we assume that the saddle point SDW order parameter 
${\bf N}$ is a real vector in ${\bf r}$-space, and that    
it has Fourier components determined by 
Fermi surface nesting vectors  
${\bf q} = {\bf Q}_i = (\pm Q_a, \pm Q_b, \pm Q_c)$~\cite{ishiguro-98}. 
In this case, the coefficients $B({\bf Q}_i)$ are identical for all 
${\bf Q}_i$'s, since the lattice dispersion in invariant under
reflections and inversions compatible with the $D_{\rm 2h}$ group.
In addition, the coefficients of all higher order terms involving 
${\bf N}({\bf Q}_i)'s$ share the same properties.
Given that ${\bf N} ({\bf r})$ is real, 
and that we have periodic boundary conditions, 
we can choose a specific reference phase where 
${\bf N}({\bf Q}_i)$ are real and identical. 
Thus, we define ${\bf N}_0 \equiv {\bf N}({\bf Q}_i)$ 
for all $i$, and the quadratic terms are dominated in the long
wavelength limit by  
$S_2^{\rm TSC} \approx A(0) \vert {\bf D}_0 \vert^2 $ 
and $S_2^{\rm SDW} \approx (m/2) B({\bf Q}_1) \vert {\bf N}_0 \vert^2$, 
respectively. Here, $m$ is the number of nesting vectors, 
and ${\bf Q}_1 = (Q_a, Q_b, Q_c)$ is chosen for definiteness.

Notice that the two order parameters 
${\bf D}({\bf p})$ and ${\bf N}({\bf q})$ 
do not couple to quadratic order,
because TSC and SDW are instabilities in particle-particle and 
particle-hole channels, respectively. 
Thus, the two orders are independent to this order, 
and their corresponding vector order parameters are free to rotate. 
However, this freedom is lost when fourth order terms are included. 

The coupling between ${\bf D}$ and ${\bf N}$ in fourth order is given by
\begin{equation}
\label{eqn:4-coupling}
S_4^{\rm C} = (C_1 + C_2/2) \vert {\bf D}_0 \vert^2 
\vert {\bf N}_0 \vert^2 
- 
C_2 \vert {\bf D}_0 \cdot {\bf N}_0 \vert^2,
\end{equation}
where
$C_1 = m T \sum_{k} G({\bf k},\omega_n) 
G^2(-{\bf k},-\omega_n) G(-{\bf k}+{\bf Q}_1, -\omega_n)
\phi_\Gamma(-{\bf k}) \phi_\Gamma({\bf k})$,
and the other coefficient
$C_2 = m T \sum_{k} G({\bf k},\omega_n) 
G({\bf k}+{\bf Q}_1, \omega_n) G(-{\bf k}-{\bf Q}_1, \omega_n) 
G(-{\bf k},-\omega_n) \phi_\Gamma(-{\bf k}-{\bf Q}_1) \phi_\Gamma({\bf k})$.
Notice that the second term on Eq.~(\ref{eqn:4-coupling}) can 
be parametrized as 
$-C_2 \cos^2(\theta) \vert {\bf D}_0 \vert^2 \vert {\bf N}_0 \vert^2$, 
where 
$\cos^2\theta \equiv \vert {\bf D}_0 \cdot {\bf N}_0 \vert^2/
\vert {\bf D}_0 \vert^2 \vert {\bf N}_0 \vert^2 \le 1$ is independent
of $\vert {\bf D}_0 \vert$ and  $\vert {\bf N}_0 \vert$.
Since ${\bf D}_0$ is unitary, its global phase can be eliminated in 
$S_4^{\rm C}$, and $\theta$ can be regarded as the angle between 
${\bf D}_0$ and ${\bf N}_0$. The coefficient $C_2$ for ${\rm (TMTSF)_2 PF_6}$ 
is positive, indicating that ${\bf D}_0$ and ${\bf N}_0$ tend
to be aligned ($\theta = 0$) or anti-aligned ($\theta = \pi$).

\begin{figure}
\begin{center}
\psfrag{mB(Q)/2}{$\frac{mB({\bf Q}_1)}{2}$}
\psfrag{A(0)}{$A(0)$}
\psfrag{P}{$P$}
\psfrag{T}{$T$}
\psfrag{TSDW}{$T_{\rm SDW}$}
\psfrag{TTSC}{$T_{\rm TSC}$}
\psfrag{T1}{$T_1$}
\psfrag{T2}{$T_2$}
\includegraphics[width=8.0cm]{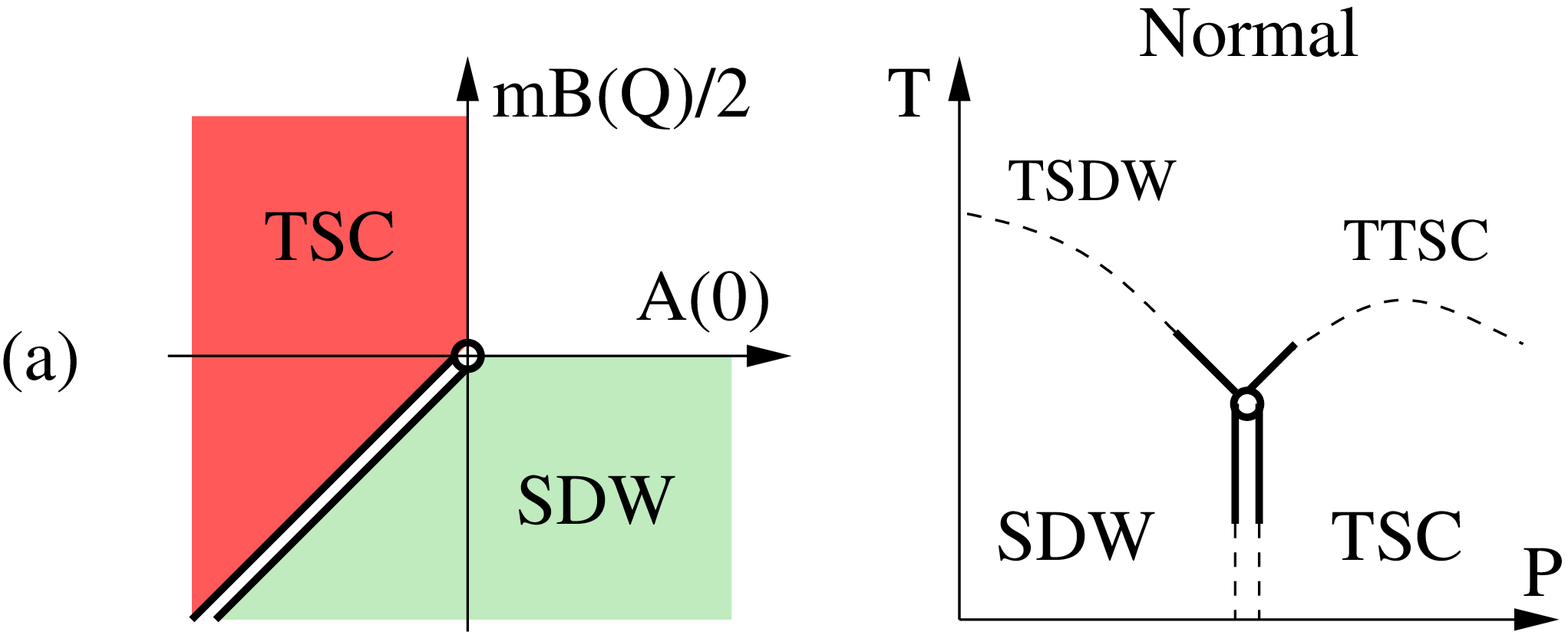}
\vskip 4mm
\includegraphics[width=8.0cm]{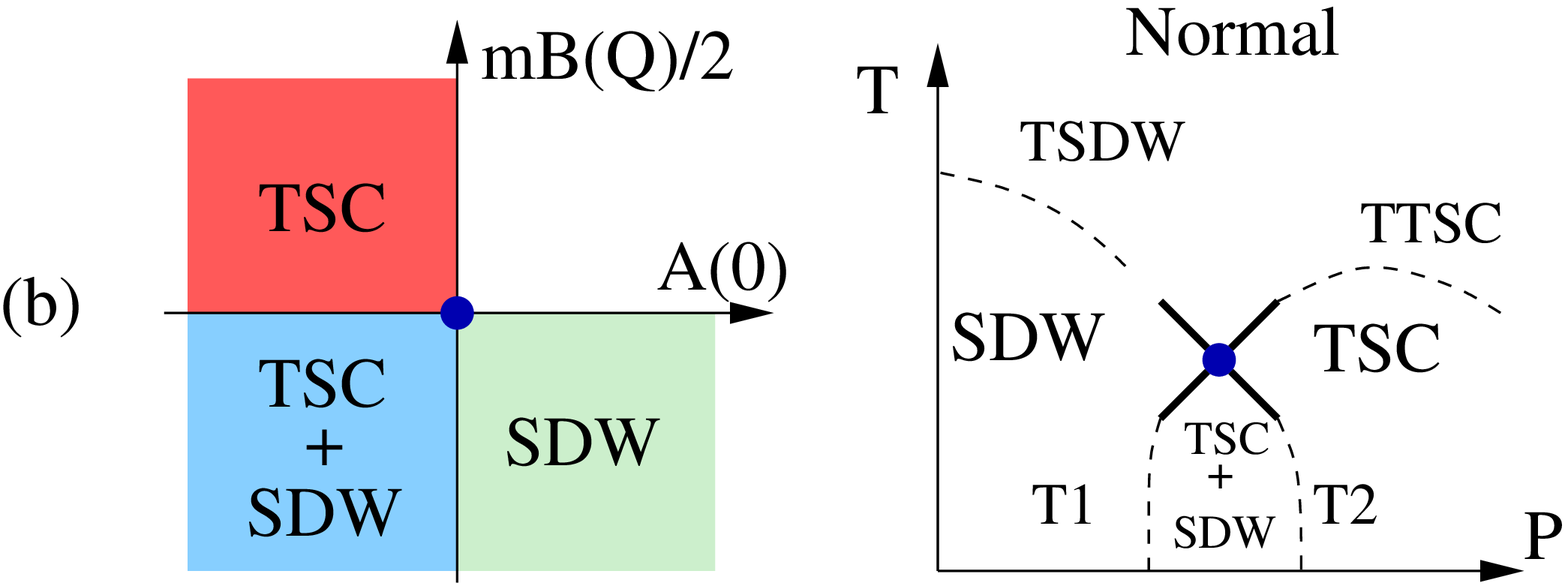}
\end{center}
\caption{$P$-$T$ phase diagrams indicating 
(a) first order transition line with no coexistence phase, 
and (b) two second order lines with coexistence region 
between TSC and SDW phases.}
\label{fig:phase}
\end{figure}

Additional fourth order terms are
\begin{equation}
\label{eqn:4-TSC-SDW}
S_4^{\rm TSC} = D_1 \vert {\bf D}_0\vert^4; \quad
S_4^{\rm SDW} = D_2 \vert {\bf N}_0 \vert^4,
\end{equation}
where $D_1 = T \sum_{k} G^2({\bf k},\omega_n) 
G^2(-{\bf k},-\omega_n) \phi_\Gamma^2({\bf k}) \phi_\Gamma^2(-{\bf k})$, 
and $D_2 = (m/2)T \sum_{k} G({\bf k},\omega_n) 
G^2({\bf k}+{\bf Q}_1, \omega_n) 
[G({\bf k},\omega_n) + G({\bf k}+2{\bf Q}_1,\omega_n)] 
+ 
(m/4) T \sum_{k} G({\bf k},\omega_n)G({\bf k}+{\bf Q}_1,\omega_n) 
\sum_{i}^\prime G({\bf k}+{\bf Q}_1+{\bf Q}_i,\omega_n)
[G({\bf k}+{\bf Q}_i,\omega_n)+ G({\bf k}+{\bf Q}_1,\omega_n)]$. 
Here $\sum_i^\prime$ represents $\sum_{{\bf Q}_i \neq \pm {\bf Q}_1}$.
This leads to the effective action
\begin{equation}
\label{eqn:action}
S_{\rm eff} = S_0 + S_2 + S_4,
\end{equation}
where $S_0$ is the normal state contribution, 
$S_2 = S_2^{\rm TSC} + S_2^{\rm SDW}$, 
and $S_4 = D_1 \vert {\bf D}_0 \vert^4 + D_2 \vert {\bf N}_0 \vert^4 + 
C(\theta) \vert {\bf D}_0 \vert^2 \vert {\bf N}_0 \vert^2$ with 
$C(\theta) = C_1+C_2/2 -C_2 \cos^2 \theta$.
The phase diagram that emerges from this action leads to either
bicritical or tetracritical points as illustrated in Fig. \ref{fig:phase}.
When $ R = C^2 (0)/(4 D_1 D_2) > 1$ the critical point $(P_c, T_c)$ 
is bicritical and there is a first order transition line 
at $(m/2) B ({\bf Q}_1) = A (0)$ when both $ B ({\bf Q}_1) < 0$ 
and $A (0) < 0$, as seen in Fig. \ref{fig:phase}(a).
However when $R < 1$, $(P_c, T_c)$ is tetracritical 
and a coexistence region for TSC and SDW orders occurs when 
both $ B ({\bf Q}_1) < 0$ and $A (0) < 0$, as seen in Fig. \ref{fig:phase}(b).
The action $S_{\rm eff}$ obtained in three dimensions 
is not SO(4) invariant, and SO(4) symmetry based theories~\cite{podolsky-04} 
can only be applied to one-dimensional systems, 
but not to the highly anisotropic but three-dimensional Bechgaard salts.

The ratio $R \approx (0.12)^2 < 1$ for 
the Bechgaard salt ${\rm (TMTSF)_2 PF_6}$ around $(P_c, T_c)$, 
when the interaction strengths $V$, $J$ 
are chosen to give the same $T_c = 1.2$ K 
at quarter filling for parameters $\vert t_x \vert = 5800$ K, 
$\vert t_y \vert =1226$ K, $\vert t_z \vert = 58$ K, used in combination
with $\phi_\Gamma ({\bf k}) = \sin(k_x a)$ ($p_x$-symmetry for TSC)
and the nesting vectors ${\bf Q} = (\pm \pi/2a, \pm \pi/2b, 0)$ ($m=4$).
This shows that ${\rm (TMTSF)_2 PF_6}$ has an 
TSC/SDW coexistence region as suggested by 
experiments~\cite{vuletic-02, kornilov-04, lee-05}. 

\begin{figure}
\begin{center}
\psfrag{x}{(b)}
\psfrag{y}{(a)}
\psfrag{b}{}
\psfrag{c}{}
\includegraphics[width=4.0cm]{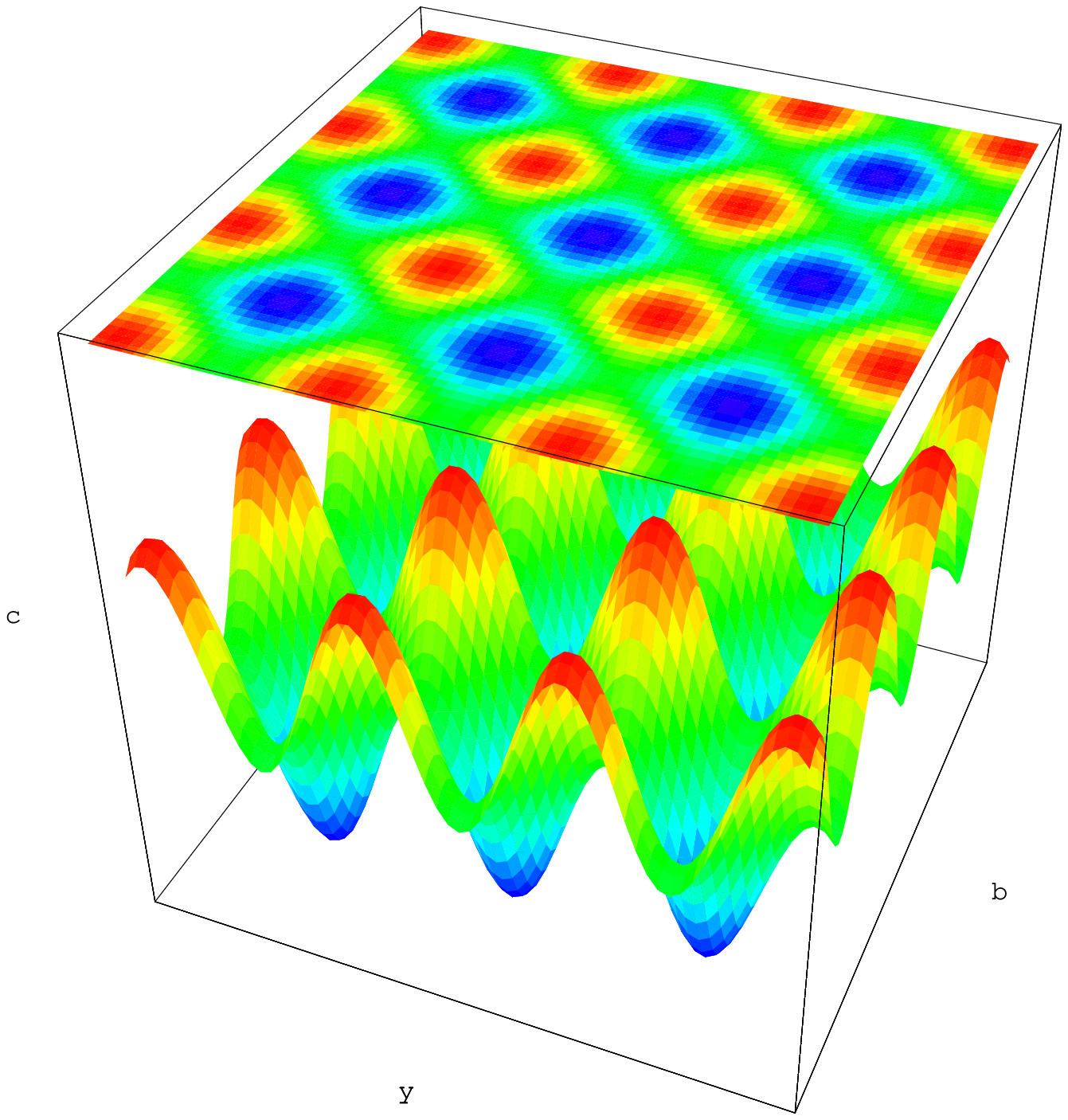}
\includegraphics[width=4.0cm]{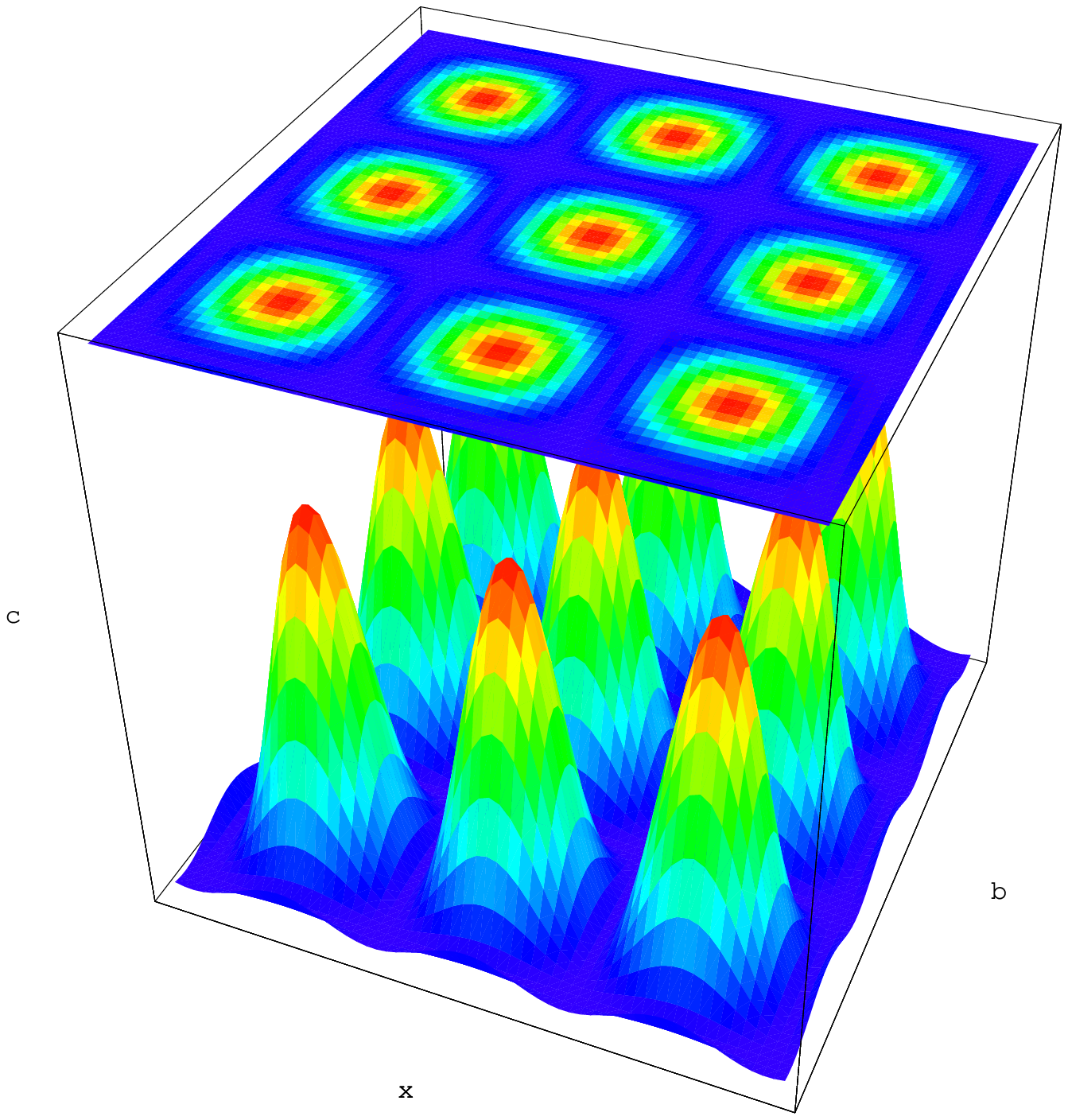}
\end{center}
\caption{Magnitude square of (a) TSC and (b) SDW order parameters 
in the coexistence region.}
\label{fig:modulation}
\end{figure}

To investigate the TSC/SDW coexistence region 
the effective action (\ref{eqn:action}) 
[with ${\bf Q} = (\pm Q_a, \pm Q_b, 0)$] is Fourier transformed into 
real space to give the Ginzburg--Laudau (GL) free energy density 
\begin{equation}
\label{eqn:GLenergy}
{\cal F} = {\cal F}_n + {\cal F}_{\rm TSC} + {\cal F}_{\rm SDW} 
+ {\cal F}_{\rm C},
\end{equation}
where ${\cal F}_n$ is the normal state contribution, 
and ${\cal F}_{\rm C} = C(\theta) \vert {\bf N}({\bf r}) \vert^2 
\vert {\bf D}({\bf r}) \vert^2 $ is the coupling term 
of the two order parameters. 
The TSC contribution is 
${\cal F}_{\rm TSC} = A(0) \vert {\bf D}({\bf r})\vert^2 
+ D_1 \vert  {\bf D}({\bf r})\vert^4 
+ \sum_{ij}\gamma_{\rm TSC}^{ij} [\partial_i {\bf D}({\bf r})] 
\cdot [\partial_j {\bf D}({\bf r})]$, 
where $\gamma_{\rm TSC}^{ij}$ is obtained from a small ${\bf p}$ expansion 
of $A({\bf p})$. The SDW contribution is  
${\cal F}_{\rm SDW} = \int d{\bf r}^\prime \left[ B ( {\bf r}, {\bf r}^{\prime} )
{\bf N} ({\bf r})\cdot {\bf N} ({\bf r}^{\prime})\right] 
+ (D_2/m^2) \vert {\bf N} ({\bf r}) \vert^4 $,
where $B ( {\bf r}, {\bf r}^{\prime} )$ is the Fourier transform of $B ({\bf q})$
in Eq.~(\ref{eqn:2-order}). For the Bechgaard salt parameters, 
the prefactor $C (0)$ of 
the coupling term ${\cal F}_{\rm C}$ is positive, and hence represents a local 
repulsive interaction between the TSC and SDW order parameters. 
As a consequence, the TSC order parameter is non-uniform
in the TSC/SDW coexistence region, and has a modulation induced 
by the SDW order parameter. 
Since $R \ll 1$ for ${\rm (TMTSF)_2 PF_6}$, the 
coupling term ${\cal F}_{\rm C}$ is small in comparison with the 
other fourth order coefficients $D_1, D_2$, and a perturbative solution is possible 
for $\vert {\bf D} ({\bf r})\vert $ and $ \vert {\bf N} ({\bf r})\vert $.
At assumed zero TSC/SDW coupling $C(0) = 0$,
the saddle point modulation for the SDW order parameter is 
${\bf N} ({\bf r}) =  m {\bf N}_0 \cos ({\bf Q}_1 \cdot {\bf r})$, with
$\vert {\bf N}_0 \vert = [- mB({\bf Q}_1)/3 D_2 ]^{1/2}$, 
while the saddle point for the TSC order parameter 
is ${\bf D} ({\bf r}) =  {\bf D}_0$ with
$\vert {\bf D}_0 \vert = [-A(0)/2 D_1]^{1/2}$. 
Including the coupling ${\cal F}_{\rm C}$ the new solution 
for the magnitude of TSC order parameter is  
\begin{eqnarray}
\label{eqn:firstorder}
\vert {\bf D}({\bf r}) \vert - \vert {\bf D}_0 \vert &=& 
-v \frac{\vert N_0 \vert^2}{\vert D_0 \vert} R^{1/2}
\Big[
\frac{\cos(2 Q_a x)}{4+ 8\xi_{x}^2 Q_a^2} 
+ \frac{\cos(2 Q_b y)}{4+ 8\xi_{y}^2 Q_b^2} 
\nonumber \\ 
&& \hspace{-5mm}
+ \frac{\cos(2 Q_a x) \cos(2 Q_b y)}
{4 + 8\xi_{x}^2 Q_a^2 + 8 \xi_{y}^2 Q_b^2 }
+ \frac{1}{4}
\Big],
\end{eqnarray}
which shows explicitly $2Q_a$ and $2Q_b$ modulations 
along the ${\bf a}$ and ${\bf b}^\prime$ axes, respectively.
Here, $\xi_{i}  = \sqrt{\vert \gamma_{\rm TSC}^{ii}/A(0) \vert}$ 
represents the TSC coherence length along the $i$ direction, 
and $v = (6 D_2 / D_1)^{1/2}$. 
Notice that the modulation in $\vert {\bf D}({\bf r}) \vert $
disappears as the SDW order goes away $\vert {\bf N}_0 \vert \to 0$.
The qualitative behavior of $\vert {\bf D}({\bf r}) \vert$ is
shown in Fig.~\ref{fig:modulation}(a).
The new solution for the SDW order parameter to the first order 
correction is $\vert {\bf N} ({\bf r}) \vert = 
\sum_i (1-R^{1/2} \vert {\bf D}_0 \vert^2/4 v \vert {\bf N}_0\vert^2 )
\vert {\bf N}_0 \vert  \cos ({\bf Q}_i \cdot {\bf r})$, 
and can be seen in Fig.~\ref{fig:modulation}(b).
Notice that the maxima of $\vert {\bf D} ({\bf r}) \vert$ 
coincide with the minima of $\vert {\bf N} ({\bf r}) \vert$
indicating that the TSC and SDW orders try to be locally excluded. 
Since the TSC and SDW modulations are out of phase, experiments that are
sensitive to the spatial distribution of the spin density or Cooper pair
charge density may reveal the coexistence of these inhomogeneous phases.

Next, we analyze the effect of magnetic fields 
on this coexistence region.
A uniform magnetic field ${\bf H}$ couples with charge via 
the Peierls substitution ${\bf k} \to {\bf k} - \vert e \vert {\bf A}$ in the 
dispersion relation given in Eq. (\ref{eqn:dispersion}), 
where ${\bf A}$ is the vector potential, and couples 
with spin via the paramagnetic term 
${\cal H}_{\rm P} = - \mu_0 {\bf H} \cdot \sum_{{\bf k},\alpha\beta} 
c_{{\bf k}\alpha}^\dagger \textrm{\boldmath $\sigma$}_{\alpha\beta} 
c_{{\bf k}\beta}$, where $\mu_0$ is the effective magnetic moment.
Thus, the effective Hamiltonian becomes
\begin{equation}
\label{eqn:MFH2}
{\cal H}_{\rm eff} = {\cal H}_0({\bf k} \to {\bf k}- \vert e \vert {\bf A})
+ {\cal H}_{\rm TSC} + {\cal H}_{\rm SDW} + {\cal H}_{\rm P}.
\end{equation}
Upon integration of the fermions, the corresponding effective action 
is 
\begin{equation}
\label{eqn:action2}
S_{\rm eff}({\bf H}) = S_0({\bf H}) + 
S_2({\bf H}) + S_4({\bf H}),
\end{equation}
where $S_0({\bf H}) = S_0 + \vert {\bf H} \vert^2 /8\pi 
- \chi_n \vert {\bf H}\vert^2/2$, $\chi_n$ is the uniform 
electronic spin susceptibility of the normal state, 
$S_2({\bf H})$ is obtained from $S_2$ by the Peierls substitution, 
and $S_4({\bf H}) = S_4 + (E_1 + E_2/2) \vert {\bf H} \vert^2 
\vert {\bf D}_0 \vert^2 - E_2 \vert {\bf H} \cdot {\bf D}_0 \vert^2 
+
(F_1 - F_2/2)\vert {\bf H} \vert^2 
\vert {\bf N}_0 \vert^2 + F_2 \vert {\bf H} \cdot {\bf N}_0 \vert^2$.
The coefficients are
$E_1=2 \mu_0^2 T \sum_{k} 
G^3(-{\bf k}, \omega_n) G({\bf k},-\omega_n) 
\phi_\Gamma({\bf k}) \phi_\Gamma(-{\bf k})$, 
$E_2 = 2 \mu_0^2 T \sum_{k}
G^2({\bf k}, \omega_n) G^2(-{\bf k},-\omega_n) 
\phi_\Gamma({\bf k}) \phi_\Gamma(-{\bf k})$, 
$F_1 = m \mu_0^2 T \sum_{k} 
G^3({\bf k},\omega_n) 
[G({\bf k}+{\bf Q}_1,\omega_n) + G({\bf k}-{\bf Q}_1,\omega_n)]$, 
and $F_2 = m \mu_0^2 T \sum_{k} G^2({\bf k},\omega_n) 
G^2({\bf k}+{\bf Q}_1, \omega_n)$. 
A detailed calculation shows that the coefficient $ E_1 = - E_2/2$, 
hence the coupling of ${\bf H}$ to ${\bf D}$ 
can be described in the more familiar form 
$F_M - \sum_{\mu\nu} H_\mu \chi_{\mu\nu} H_\nu /2 $, 
where $\chi_{\mu\nu} = \chi_n \delta_{\mu\nu} + E_2 D_\mu^* D_\nu$.

For Bechgaard salts, the coefficients $E_2<0$ and $F_2>0$ indicating that
${\bf D}$ and ${\bf N}$ prefer to be perpendicular to the magnetic
field ${\bf H}$. These conditions, when combined with $C_2 > 0$ 
in Eq. (\ref{eqn:4-coupling}), 
indicate that ${\bf D}$ and ${\bf N}$ prefer to be parallel to each 
other, but perpendicular to ${\bf H}$. 
However, the relative orientation of these
vectors in small fields is affected by spin anisotropy effects 
which were already observed 
in ${\rm (TMTSF)_2 PF_6}$, where the easy axis for ${\bf N}$  
is the ${\bf b}^\prime$ direction~\cite{mortensen-81}. 
Such an anisotropy effect can be described by 
adding a quadratic term $-u_N N_{b^\prime}^2$ with $u_N >0$, 
which favors ${\bf N} \parallel {\bf b}^\prime$. 
Similarly, the ${\bf D}$ vector also has anisotropic effect 
caused by spin-orbit coupling, 
and can be described by adding a quadratic term $-u_D D_{i}^2$, 
where $i$ is the easy axis for TSC. 
(Quartic TSC and SDW terms also become anisotropic.)

However, a sufficiently large ${\bf H}\parallel {\bf b}^\prime$ 
can overcome spin anisotropy effects, and drive the 
${\bf N}$ vector to flop onto the ${\bf a}$-${\bf c}^*$ plane.
This canting (flop) transition was reported~\cite{mortensen-81}
in ${\rm (TMTSF)_2PF_6}$ for $H \approx 1\ {\rm T}$ 
at zero pressure and $T = 8\ {\rm K}$. 
\begin{figure}
\begin{center}
\psfrag{H1}{$H_1$}
\psfrag{H2}{$H_2$}
\psfrag{T1}{$T_1$}
\psfrag{T2}{$T_2$}
\psfrag{TSDW}{$T_{\rm SDW}$}
\psfrag{TTSC}{$T_{\rm TSC}$}
\psfrag{HF}{$H_F$}
\psfrag{HSDW}{$H_{\rm SDW}$}
\psfrag{P<Pc}{$P<P_c$}
\psfrag{P>Pc}{$P>Pc$}
\psfrag{T}{$T$}
\psfrag{SDW}{SDW}
\psfrag{TSC}{TSC}
\psfrag{H//b'}{${\bf H} \parallel {\bf b}^\prime$}
\psfrag{(a)}{(a)}
\psfrag{(b)}{(b)}
\includegraphics[width=8.5cm]{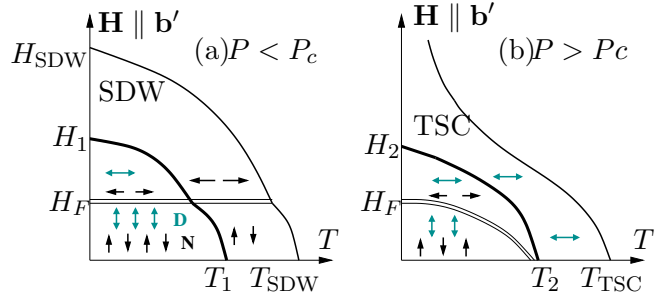}
\end{center}
\caption{$H$-$T$ phase diagrams showing the TSC/SDW coexistence region (thick solid line) 
and canting transitions (double line) for (a) $P<P_c$ and (b) $P>P_c$. }
\label{fig:magnetic}
\end{figure}
If such a spin-flop transition persists 
near the TSC/SDW critical point ($P_c, T_c$)
as suggested in our discussion,
then the flop transition of the ${\bf N}$ vector forces 
the ${\bf D}$ vector to flop as well, 
and has potentially serious consequences to the superconducting state. 
Schematic phase diagrams are are shown in Fig.~3(a) and 3(b). 
For $P < P_c$, if a flop transition occurs for $H_F < H_1 (0)$ [see Fig. 3(a)], 
then ${\bf N}$ flops both in the pure SDW and in the
TSC/SDW coexistence phases, in which case it forces ${\bf D}$ vector to flop as well. 
If the flop transition occurs for $H_{\rm SDW} (0) < H_F < H_1 (0)$ (not shown)
then only the pure SDW phase is affected.
This situation is qualitatively different for $P > P_c$.
In the zero (weak) spin-orbit coupling limit the ${\bf D}$ vector 
is free to rotate in a magnetic field and tends to be perpendicular
to ${\bf H}$ in order to minimize its magnetic free energy $F_M$. 
Thus, for ${\bf H} \parallel {\bf b}^\prime$ and $\vert {\bf H} \vert > H_2$, 
the ${\bf D}$ vector lies in the ${\bf a}$-${\bf c}^*$ plane since there is no SDW order. 
However, at lower temperatures and small magnetic fields when TSC and SDW orders coexist, 
the spin anisotropy field forces ${\bf N}$ to be along ${\bf b}^r\prime$ and 
${\bf N}$ forces ${\bf D}$ to flop from the ${\bf a}$-${\bf c}^*$ plane to ${\bf b}^\prime$ direction.
This canting transition occurs at $H_F < H_2 (0)$ [see Fig. 3(b)], 
when ${\bf N}$ flops in the TSC/SDW coexistence phase, and forces the 
${\bf D}$ vector to flop as well.

In summary, we showed that the TSC and SDW order parameters can coexist in 
the $P$-$T$ phase diagram of quasi-one-dimensional organic conductors. 
In the coexistence region the TSC order parameter is non-uniform, and its modulation 
is induced via the SDW order parameter.
We also showed that theories based on SO(4) symmetry cannot be applied to 
these highly anisotropic three-dimensional systems, 
since they are strictly valid only in the one-dimensional limit. 
Furthermore, we discussed qualitatively magnetic field effects on 
the coexistence region. We proposed that a magnetic field induced canting transition 
of the SDW order parameter 
affects dramatically the phase diagram of the coexistence region, 
both below and above the critical pressure. 
We would like to thank NSF for support (DMR-0304380).

\end{document}